\documentclass[pdflatex]{sn-jnl}

\usepackage{graphicx,alltt}
\usepackage{amsmath,amssymb,amsfonts}
\usepackage[T1]{fontenc}
\usepackage[utf8]{inputenc}
\usepackage{isabelle,isabellesym}
\renewenvironment{isabelle}
{\begin{trivlist}\vskip1ex\begin{isabellebody}\item\relax}
{\end{isabellebody}\vskip1ex\end{trivlist}}

\usepackage{amsthm}

\let\ts=\thinspace
\let\ge=\geqslant
\let\le=\leqslant
\newcommand{\card}[1]{\lvert#1\rvert}
\newtheorem*{theorem}{Theorem}
\newtheorem*{lemma}{Lemma}

\begin{document}
\title[Formalising New Mathematics]{Formalising New Mathematics in Isabelle: Diagonal Ramsey}
\author{\fnm{Lawrence} \sur{Paulson}}\email{lp15@cam.ac.uk}

\affil{\orgdiv{Computer Laboratory}, \orgname{University of Cambridge}, \\\orgaddress{\street{15 JJ Thompson Avenue}, \city{Cambridge}, \postcode{CB3 0FD}, \country{UK}}}

\abstract{The formalisation of mathematics is starting to become routine, 
but the value of this technology to the work of mathematicians remains to be shown. 
There are few examples of using proof assistants to verify brand-new work. 
This paper reports the formalisation of a major new result~\cite{campos-exponential-ramsey} 
about Ramsey numbers that was announced in 2023. 
One unexpected finding was the heavy need for computer algebra techniques.}


\maketitle

\section{Introduction}

In the summer of 2017, a six week programme entitled \textit{Big Proof} was held at Cambridge's Isaac Newton Institute,
``directed at the challenges of bringing proof technology into mainstream mathematical practice''.
It brought together the leading figures in the formalisation of mathematics from around the world,
and marked the start of the present era of large libraries of formalised mathematics.
It is now clear that material from almost any branch of mathematics can be formalised,
raising the question of whether further work of this sort serves any scientific purpose. 
But the purpose of interactive proof assistants, from their origin decades ago, 
has been to \textit{assist} in the \textit{construction} of proofs. The clue is in the name.

Until 2017, we could only point to a few instances where machine-assisted proof had benefited mathematicians:
\begin{itemize}
	\item McCune showed by an automatic proof that every Robbins algebra was boolean~\cite{mccune-solution}.
	\item Gonthier's formalisation of the Four Colour Theorem~\cite{gonthier-4ct} settled doubts about the 1976 Appel--Haken proof, 
    	which heavily relied on computer calculations.
    \item Hales's proof of the Kepler conjecture similarly relied on checking cases by computer. After a years-long effort, the proof was  verified and indeed simplified \cite{hales-revision-proof}. 
    \item My own formalisation of Gödel's second incompleteness theorem~\cite{paulson-incompl-logic} was motivated by a comment 
       from a prominent logician (at an earlier Newton Institute programme) about the obscurity of its proofs.
\end{itemize}
Although these examples were not enough to convince mathematicians that 
proof assistants were ready to tackle contemporary material,
now we are starting to see mathematical results verified prior to journal publication.
Finally, proof assistants have checked deep, original results before any human referees:
\begin{itemize}
	\item The \textit{Liquid Tensor Experiment}: an effort to verify a lemma by Scholze and Clausen in the new field of \textit{condensed mathematics}. 
	They were unsure about their large and complex proof, appealed to the research community, and the key result was formally checked within six months.
	\item A breakthrough on Ramsey numbers: an exponential reduction in the upper bound for Ramsey numbers~\cite{campos-exponential-ramsey}, 
	announced in March 2023. A Cambridge PhD student, Bhavik Mehta, formalised this sensational result within five months.%
	\footnote{\url{https://xenaproject.wordpress.com/2023/11/04/formalising-modern-research-mathematics-in-real-time/}}
\end{itemize}
Both of these were done using Lean, a proof assistant based on a dependent type theory.
That raised the question of whether Isabelle, a leading proof assistant based on simple type theory, could also be suitable for such proofs.
Condensed mathematics was well beyond my ability, so I decided to tackle Ramsey.
I would work from the original article, and I would peek at Mehta's proof when stuck (details of that below).

Isabelle and Lean are proof assistants in which types, functions and other entities can be defined 
and theorems proved about them.
Descriptions and documentation are widely available elsewhere 
(e.g. for Isabelle \cite{nipkow-isar-tutorial} \cite[\S3]{paulson-formalised-theorem}).
The chief difference between Isabelle and Lean is that types in the latter can depend on arbitrary parameters
and not merely upon other types. The advantage of Isabelle's less expressive formalism is better automation
and the avoidance of dependency issues: 
chiefly, that $i=j$ does not imply that the types $T(i)$ and $T(j)$ are the same.
The Ramsey proof has nothing suggestive of dependent types, but it relies on a surprising diversity of automation methods.
The point of the exercise is to test how well Isabelle copes with 
the formalisation of a large, technical piece of research mathematics.

The rest of the paper is as follows: An introduction to Ramsey's theorem~(\S2) 
is followed by a sketch of the new algorithm reported in the Ramsey paper~(\S3).
In the next three sections this algorithm is formalised~(\S4), 
one proof is explored in detail~(\S5) and the rest of the construction is sketched~(\S6).
There follows a discussion of the computer algebra aspects of the proof~(\S7)
and finally, discussion and conclusions~(\S8).

\section{Ramsey's theorem}

In its simplest form, Ramsey's theorem is concerned with graphs.
We imagine the edges to be coloured and are interested in monochromatic (single-coloured) complete subgraphs, 
or \textit{cliques}. A $k$-clique is a clique of $k$ vertices;
an $n$-graph is a graph of $n$ vertices.

\begin{theorem}[Ramsey]
For all $k$ and $\ell$ there exists a number $R(k,\ell)$ such that
if the edges of a complete $R(k,\ell)$-graph $G$ are coloured red or blue,
then $G$ will contain either a red $k$-clique or a blue $\ell$-clique.
\end{theorem}

\begin{figure}[hbt]
\begin{center}
\includegraphics[width=120pt]{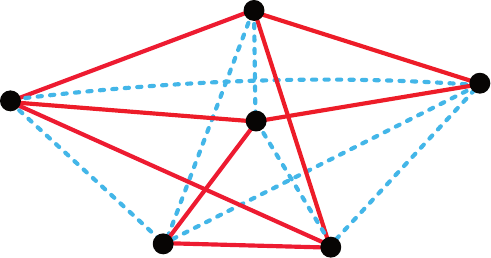}	
\end{center}
\caption{An example to illustrate $R(3,3)=6$. Solid lines are red. Two cliques can be seen.}\label{fig:ramsey}	
\end{figure}

Abbreviate $R(k,k)$ as $R(k)$, for Ramsey numbers on the \textit{diagonal}.
Figure~\ref{fig:ramsey} shows a graph of size 6 containing a monochromatic clique of size~3,
to illustrate that $R(3)=6$. It's known that $R(4)=18$ and that $43\le R(5)\le 46$.
Erdős joked about the difficulty of determining $R(5)$ more than 30 years ago, and we still don't know it.
The claim that $R(5)\le 46$ is extremely recent and as yet unconfirmed~\cite{r55le46}.

This theorem can be generalised to possibly infinite hypergraphs and much else. 
One focus of Ramsey theory is to establish lower and upper bounds for Ramsey numbers.
Already in 1935, Erdős and Szekeres found an upper bound; Erdős later proved a lower bound, giving
$$ \sqrt2^k \le R(k) \le {2k-2 \choose k-1}< 4^k. $$
These bounds are wide apart, and subsequent refinements over decades failed to improve the bases of the exponents:
hence the enthusiasm when Campos et al.\ announced a proof that $R(k)\le(4-\epsilon)^k$, 
where $\epsilon$ was small but positive.

There is a short proof of Ramsey's theorem that establishes the $4^k$ bound:
in fact, that $R(k,\ell) < 2^{k+\ell}$.
It involves the execution of a simple non-deterministic algorithm operating on sets of vertices $X$, $A$, $B$. 
First, some notation: $N_R(x)$ denotes the \textit{red neighbours} of $x$ 
(the set of all $y$ connected to $x$ by a red edge).
Similarly, $N_B(x)$ is the \textit{blue neighbours} of~$x$.

Begin by putting all the vertices of the graph into $X$.
Then, repeat the following as long as $X$ is nonempty and $\card A < k$, $\card B <\ell$:
\begin{enumerate}
	\item Pick an arbitrary vertex $x\in X$.
	\item If $x$ has no fewer red neighbours than blue in $X$, then set \\ $X\to N_R(x)\cap X$ and $A\to A\cup \{x\}$ 
	\item Otherwise, set \\ $X\to N_B(x)\cap X$ and $B\to B\cup \{x\}$.
\end{enumerate}
Step~2 leaves in $X$ only the red neighbours of~$x$.
Because the same had been done for all the elements previously added to~$A$, 
the edges between $x$ and elements of~$A$ are all red, and by induction, $A$ is a red clique.
Analogous remarks hold for step~3:
the algorithm builds a red clique in $A$ and a blue clique in~$B$.
After at most $k+\ell-1$ steps, we must have either a red $k$-clique or a blue $\ell$-clique.
Because each iteration (step~2 or step~3) removes at most half the elements of $X$, 
a clique will be found if initially $\card X \ge 2^{k+\ell}$.

\begin{figure}[hbt]
\begin{center}
\includegraphics[width=250pt]{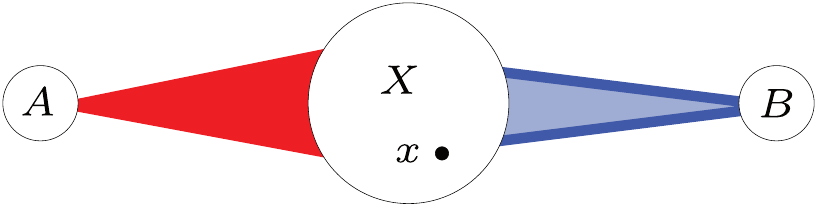}	
\end{center}
\caption{An algorithm for proving Ramsey's theorem. (The hollow colour is blue.)}
\label{fig:simple-book}	
\end{figure}

Figure~\ref{fig:simple-book} shows the algorithm in operation. We see the set $X$ of vertices and
the arbitrarily chosen element~$x$. The red expresses that all edges between $X$ and $A$ are red
(as are all edges within $A$); similarly, all edges between $X$ and $B$ are blue.
This property is an invariant that holds initially (because $A$ and $B$ are empty) and is preserved by steps 2 and~3.

\section{The improved Ramsey algorithm}

The algorithm of Fig.\ts\ref{fig:simple-book} is unsophisticated in that the choice of $x$ is arbitrary. 
Campos et al. \cite{campos-exponential-ramsey} get a sharper bound from a better algorithm.
They break the symmetry between red and blue, adding vertices to $A$ one at a time 
but adding huge chunks of vertices to~$B$.
And they divide the vertices of the original graph into two separate sets, $X$ and~$Y$.

Key to their algorithm (and indeed to the simpler one) is the notion of a \textit{book}:
a pair of sets $(S,T)$, where all the edges from $S\cup T$ into $S$ have the same colour 
(so in particular, $S$ is a monochromatic clique).

\begin{figure}[hbt]
\begin{center}
\includegraphics[width=250pt]{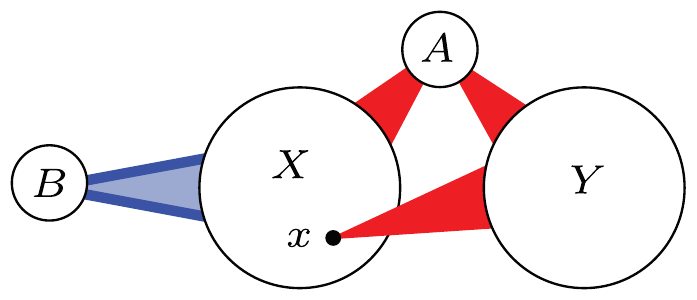}
\end{center}
\caption{The Book Algorithm. (The hollow colour is blue.)}
\label{fig:book-algo}	
\end{figure}

The setup for their \textit{book algorithm} includes integers $k$, $\ell$, with $\ell\le k$.
A complete $n$-graph $G$ is coloured red/blue with no red $k$-clique or blue $\ell$-clique.
The algorithm manipulates the disjoint sets $X$, $Y$, $A$, $B$, where initially $A=B=\emptyset$
while $X$ and $Y$ are an arbitrary partition of the vertices of~$G$ into equal ($\pm1$) parts.
A further parameter is the constant $\mu\in(0,1)$.
Figure~\ref{fig:book-algo} also shows some $x\in X$ being chosen on the basis of the density
of red edges linking its neighbours with $Y$.
(The \textit{density} of edges between $X$ and $Y$ equals the number of such edges divided by $\card X\card Y$.)

The process runs until either $\card X \le R(k, \ell^{\lceil 3/4\rceil})$ or $p\le 1/k$,
where $p$ denotes the density of red edges between $X$ and~$Y$.
Here is a sketch of the possible execution steps:
\begin{itemize}
	\item \textit{Degree regularisation} involves removing from $X$ all vertices having ``relatively few''
	      red neighbours in~$Y$. This is performed at every even-numbered execution step.
	\item An odd-numbered step is taken as follows:
	\begin{itemize}
	\item A \textit{big blue step}, if there exist $R(k, \ell^{\lceil 2/3\rceil})$ vertices
	      $x\in X$ such that $$\card{N_B(x)\cap X}\ge \mu \card X, $$
	      then a ``large'' blue book $(S,T)$ is chosen and the sets updated by
	      $$ X\to T, \quad Y\to Y, \quad A\to A,\quad B\to B\cup S.$$
	\item But if not, then a \textit{central vertex} $x\in X$ exists satisfying $\card{N_B(x)\cap X}\le \mu \card X$
	      and a second technical condition. 
	      If the density of red edges between $N_R(x)\cap X$ and $N_R(x)\cap Y$ is sufficiently high,
	      a \textit{red step} is performed:
	      $$ X\to N_R(x)\cap X, \quad Y\to N_R(x)\cap Y, \quad A\to A\cup\{x\},\quad B\to B.$$
	      Otherwise, somewhat as in the previous algorithm, $x$ is moved to $B$ instead, 
	      in a \textit{density-boost} step:
	      $$ X\to N_B(x)\cap X, \quad Y\to N_R(x)\cap Y, \quad A\to A,\quad B\to B\cup\{x\}.$$
    \end{itemize}
\end{itemize}

The paper \cite{campos-exponential-ramsey} presents further definitions specifying aspects of execution traces. 
Its next 20 or so pages are devoted to proving that big blue steps really are big,
density-boost steps really boost the density of red edges between $X$ and~$Y$, and providing lower bounds
for $Y$ and $X$ as execution proceeds;
the point is to show that they do not decrease too quickly as a function of execution steps of various kinds.
Next comes the so-called \textit{zigzag lemma}, which bounds the number of density-boost steps 
(which decrease both $X$ and $Y$).
It emerges through their work that one of the termination conditions, $p\le 1/k$, can never happen: 
the red density remains high enough.

The authors need a further 10 pages to obtain the headline result.
Here they typically take a candidate Ramsey number, construct an execution of the algorithm
by defining $X$, $Y$ and so forth, and use the lemmas proved previously to deduce their claims.
The authors first find exponentially improved bounds for $R(k,\ell)$ when $\ell\le k/9$ (``far from the diagonal''),
then when $\ell\le k/4$ (``closer to the diagonal'') and finally for $\ell=k$.

The proofs rely on a variety of elementary techniques for estimating real numbers, as well as asymptotic reasoning 
and finding upper/lower bounds for real functions. 
The probabilistic method pioneered by Erdős is used twice.
The Isabelle formalisation~\cite{Ramsey_Bounds-AFP} comprises some 12,500 lines, of which 1000 concern long-standing 
bounds on Ramsey numbers, the remainder being devoted to the new paper~\cite{Diagonal_Ramsey-AFP}.
The proofs are too long, difficult and intricate to present in full, but let's look at a few representative examples.

\section{Defining the book algorithm in Isabelle} \label{sec:book}

The diagonal paper can be seen as an abstract example of program verification.
The book algorithm is indexed by a simple step counter and operates on its four-component state.
The state variables at step $i$ are written $X_i$, $Y_i$, $A_i$, $B_i$, with $i=0$ denoting the initial state.

A succession of properties about the state are proved in terms of measures of the execution itself,
primarily the numbers of steps of the various kinds. 
The authors assign $\cal R$, $\cal B$, $\cal S$ and $\cal D$
to the sets of indices corresponding to an execution of 
a red step, a big blue step, a density-boost step and a degree regularisation step respectively,
and they write $t=\card{\cal R}$ and $s=\card{\cal S}$ for the number of red steps and density-boost steps in a full execution.
I found it convenient to introduce also $\card{\cal H}$ for the set of indices after the algorithm effectively halts.

The paper is concerned with the book algorithm's execution state but also its context:
a graph of $n$ vertices, the edges coloured red/blue and many other details.
We don't want to repeat this boilerplate in every lemma statement.
Luckily, Isabelle's \textit{locale} mechanism~\cite{ballarin-exploring} lets us declare these parameters and their constraints in one place, 
along with associated proofs and even dedicated syntax.
A locale encapsulates a local scope, but internally it is simply a predicate.

The locale declarations for the diagonal Ramsey development given an idea what we are dealing with.
For starters, although it only becomes evident later, one parameter of the construction is $p_0\in(0,1)$
specifying the minimum density of red edges between $X_0$ and $Y_0$. 

\begin{isabelle}
\isacommand{locale}\ P0\_min\ =\ \ \ \isanewline
\ \ \isakeyword{fixes}\ p0\_min\ ::\ real\isanewline
\ \ \isakeyword{assumes}\ p0\_min:\ "0\ <\ p0\_min"\ "p0\_min\ <\ 1"	
\end{isabelle}

Locales can be combined flexibly to build more complicated structures.
Here we combine a formalisation of finite simple graphs~\cite{Undirected_Graph_Theory-AFP}
with the previous locale, adding the assumptions that our graph is complete
and the requirement that the underlying type is infinite.

\begin{isabelle}
\isacommand{locale}\ Book\_Basis\ =\ fin\_sgraph\ +\ P0\_min\ +\isanewline
\ \ \isakeyword{assumes}\ complete:\ "E\ =\ all\_edges\ V"\isanewline
\ \ \isakeyword{assumes}\ infinite\_UNIV:\ "infinite\ (UNIV::'a\ set)"
\end{isabelle}

We further extend it with a colouring that has no red $k$-cliques or blue $\ell$-cliques.
\begin{isabelle}
\isacommand{locale}\ No\_Cliques\ =\ Book\_Basis\ +\isanewline
\ \ \isakeyword{fixes}\ Red\ Blue\ ::\ "'a\ set\ set"\isanewline
\ \ \isakeyword{assumes}\ Red\_E:\ "Red\ \isasymsubseteq \ E"\isanewline
\ \ \isakeyword{assumes}\ Blue\_def:\ "Blue\ =\ E-Red"\isanewline
\ \ \isakeyword{fixes}\ l::nat\ \ \ %
\isamarkupcmt{blue limit%
}\isanewline
\ \ \isakeyword{fixes}\ k::nat\ \ \ %
\isamarkupcmt{red limit%
}\isanewline
\ \ \isakeyword{assumes}\ l\_le\_k:\ "l\ \isasymle \ k"\ \isanewline
\ \ \isakeyword{assumes}\ no\_Red\_clique:\ "\isasymnot \ (\isasymexists K.\ size\_clique\ k\ K\ Red)"\isanewline
\ \ \isakeyword{assumes}\ no\_Blue\_clique:\ "\isasymnot \ (\isasymexists K.\ size\_clique\ l\ K\ Blue)"
\end{isabelle}

We need two separate locales for the book algorithm itself.
In the earlier proofs, it is governed by a given fixed $\mu\in(0,1)$:

\begin{isabelle}
\isacommand{locale}\ Book\ =\ Book\_Basis\ +\ No\_Cliques\ +\isanewline
\ \ \isakeyword{fixes}\ \isasymmu ::real\isanewline
\ \ \isakeyword{assumes}\ \isasymmu 01:\ "0\ <\ \isasymmu "\ "\isasymmu \ <\ 1"\isanewline
\ \ \isakeyword{fixes}\ X0\ ::\ "'a\ set"\ \isakeyword{and}\ Y0\ ::\ "'a\ set"\ \ \ \ %
\isamarkupcmt{initial values%
}\isanewline
\ \ \isakeyword{assumes}\ XY0:\ "disjnt\ X0\ Y0"\ "X0\ \isasymsubseteq \ V"\ "Y0\ \isasymsubseteq \ V"\isanewline
\ \ \isakeyword{assumes}\ density\_ge\_p0\_min:\ "gen\_density\ Red\ X0\ Y0\ \isasymge \ p0\_min"
\end{isabelle}

In the later proofs, the parameter is called $\gamma$ and is defined by $\gamma = \frac{\ell}{k+\ell}$:

\begin{isabelle}
\isacommand{locale}\ Book'\ =\ Book\_Basis\ +\ No\_Cliques\ +\isanewline
\ \ \isakeyword{fixes}\ \isasymgamma ::real\isanewline
\ \ \isakeyword{assumes}\ \isasymgamma \_def:\ "\isasymgamma \ =\ real\ l\ /\ (real\ k\ +\ real\ l)"\isanewline
\ \ \isakeyword{fixes}\ X0\ ::\ "'a\ set"\ \isakeyword{and}\ Y0\ ::\ "'a\ set"\isanewline
\ \ \isakeyword{assumes}\ XY0:\ "disjnt\ X0\ Y0"\ "X0\ \isasymsubseteq \ V"\ "Y0\ \isasymsubseteq \ V"\isanewline
\ \ \isakeyword{assumes}\ density\_ge\_p0\_min:\ "gen\_density\ Red\ X0\ Y0\ \isasymge \ p0\_min"
\end{isabelle}

Much of the formalisation takes place within locale \isa{Book}, with the parameters specified above
immediately accessible. Towards the end, while proving the headline claims, we start outside the locale.
Then we create a graph, colour its edges, choose $X_0$ and $Y_0$, 
thereby creating an instance of \isa{Book}.
These steps correspond  to the authors' remark ``we now apply the book algorithm''.

Here is a glimpse at the algorithm itself. Recall that an odd-numbered step can be a big blue step, a red step or a density-boost step.

\begin{isabelle}
\isacommand{definition}\ next\_state\ ::\ "'a\ config\ \isasymRightarrow \ 'a\ config"\ \isakeyword{where}\isanewline
\ \ "next\_state\ \isasymequiv \ \isasymlambda (X,Y,A,B).\ \isanewline
\ \ \ \ \ if\ many\_bluish\ X\ \isanewline
\ \ \ \ \ then\ let\ (S,T)\ =\ choose\_blue\_book\ (X,Y,A,B)\ in\ (T,\ Y,\ A,\ B\isasymunion S)\ \isanewline
\ \ \ \ \ else\ let\ x\ =\ choose\_central\_vx\ (X,Y,A,B)\ in\isanewline
\ \ \ \ \ \ \ \ \ \ if\ reddish\ k\ X\ Y\ (red\_density\ X\ Y)\ x\ \isanewline
\ \ \ \ \ \ \ \ \ \ then\ (Neighbours\ Red\ x\ \isasyminter \ X,\ Neighbours\ Red\ x\ \isasyminter \ Y,\ insert\ x\ A,\ B)\isanewline
\ \ \ \ \ \ \ \ \ \ else\ (Neighbours\ Blue\ x\ \isasyminter \ X,\ Neighbours\ Red\ x\ \isasyminter \ Y,\ A,\ insert\ x\ B)"
\end{isabelle}

The execution stepper itself must also take care of the even-numbered steps (degree regularisation) and the termination condition 
(when it leaves the next state unchanged):

\begin{isabelle}
\isacommand{primrec}\ stepper\ ::\ "nat\ \isasymRightarrow \ 'a\ config"\ \isakeyword{where}\isanewline
\ \ "stepper\ 0\ =\ (X0,Y0,\{\},\{\})"\isanewline
|\ "stepper\ (Suc\ n)\ =\ \isanewline
\ \ \ \ \ (let\ (X,Y,A,B)\ =\ stepper\ n\ in\ \isanewline
\ \ \ \ \ \ if\ termination\_condition\ X\ Y\ then\ (X,Y,A,B)\ \isanewline
\ \ \ \ \ \ else\ if\ even\ n\ then\ degree\_reg\ (X,Y,A,B)\ else\ next\_state\ (X,Y,A,B))"
\end{isabelle}

Functions like \isa{many\_bluish} and \isa{choose\_blue\_book} refer to details of the algorithm omitted here.
Many other definitions and basic lemmas are also omitted.

\section{A detailed peek at a proof} \label{sec:peek}

The very first fact to be proved about the algorithm is stated by the authors 
essentially as follows \cite[p.\ts11]{campos-exponential-ramsey}:

\begin{lemma}[4.1]
	Set $b=\ell^{1/4}$. If there are $R(k,\ell^{2/3})$ vertices $x\in X$ such that
	\[ \card{N_B(x)\cap X}\ge \mu \card X, \]
	then $X$ contains either a red $k$-clique, or a blue book $(S,T)$ with $S\ge b$ and $\card T\ge \mu^{\card S}\card X/2$.
\end{lemma}

Here $X$ is any set, not necessarily a state component of the book algorithm, 
though that's how the lemma will be applied.
The proof is under a page, but it presents many complications.
For starters, some trivia: it is already assumed that there are no red $k$-cliques in the given graph,
so we can already rule out the possibility of such a clique in~$X$.
And by convention, expressions like $R(k,\ell^{2/3})$ don't indicate 
whether the floor or the ceiling of $\ell^{2/3}$ is intended. 
The omission looks strange to a computer scientist.%
\footnote{In most cases, I have assumed the ceiling and have written it explicitly in the previous section.}

A more serious omission is that the theorem only holds for sufficiently large $\ell$ and all $k\ge\ell$.
This $\ell$ may depend upon $\mu$ alone; 
moreover, it must be constant as $\mu$ ranges over a closed subinterval of $(0,1)$.
The need for this becomes apparent only 19 pages later, in Lemma~9.3.
Constraints on $\ell$ (or, more usually, $k$) need to be picked up here and there in the proof.
It is quite common that a particular inequality does not hold in general, but holds for a sufficiently large~$k$.

So we need to decide how to formalise the ``sufficiently large'' condition.
The elegant approach uses the \isa{eventually} topological filter~\cite{hoelzl-filters}, 
which is available in both Isabelle/HOL and Lean.
Initially, I followed Mehta and used \isa{eventually}, 
making the various constraints local to each proof and thus hidden.
I later adopted the approach of making these constraints explicit:
to define a ``bigness predicate'' for each theorem, justifying each predicate
by proving a lemma (expressed using \isa{eventually}) 
to establish that the predicates does indeed hold for all sufficiently large $\ell$.

The predicate \isa{Big\_Blue\_4\_1}, which is for Lemma~4.1, is a conjunction of six inequalities, 
such as $l \ge (6/\mu) ^ {12/5}$.
Recall that the property may depend upon $\mu$ alone and the obtained $\ell$ must be good over a closed interval.
The corresponding ``bigness lemma'' refers to the closed interval $[\mu_0,\mu_1]$ where $0<\mu_0$.
Here there is no need to assume $\mu_1<1$, so the precondition could be simplified to $\mu\ge\mu_0$.
Uniformity in the lemma statements will be convenient however, since there will be lots of them.
The proof, some 40-odd lines, is routine and deals with the six inequalities separately.
\begin{isabelle}
\isacommand{lemma}\ Big\_Blue\_4\_1:\isanewline
\ \ \isakeyword{assumes}\ "0<\isasymmu 0"\isanewline
\ \ \isakeyword{shows}\ "\isasymforall \isactrlsup \isasyminfinity l.\ \isasymforall \isasymmu .\ \isasymmu \ \isasymin \ \{\isasymmu 0..\isasymmu 1\}\ \isasymlongrightarrow \ Big\_Blue\_4\_1\ \isasymmu \ l"
\end{isabelle}

Here is the formal statement of Lemma~4.1. 
It's easy to state, because the most technical parts of the theorem statement (\isa{many\_bluish}
and \isa{good\_blue\_book}) have already been formalised
for the specification of the book algorithm.
\begin{isabelle}
\isacommand{lemma}\ Blue\_4\_1:\isanewline
\ \ \isakeyword{assumes}\ "X\isasymsubseteq V"\ \isakeyword{and}\ manyb:\ "many\_bluish\ X"\isanewline
\ \ \ \ \isakeyword{and}\ big:\ "Big\_Blue\_4\_1\ \isasymmu \ l"\isanewline
\ \ \isakeyword{shows}\ "\isasymexists S\ T.\ good\_blue\_book\ X\ (S,T)\ \isasymand \ card\ S\ \isasymge \ l\ powr\ (1/4)"
\end{isabelle}

Before proving this it was necessary to build a basic library of Ramsey numbers, 
including the Erdős lower bound $2^{k/2} \le R(k,k)$ with its famous probabilistic argument.
The formal proof of Lemma~4.1 is some 350 lines. 
Let's look at the treatment of some of the estimates in this proof.

Early in the proof we obtain a blue $m$-clique, $U\subseteq X$, where $m=\ell^{2/3}$.
The text continues:

\begin{quote}
Let $\sigma$ be the density of blue edges between $U$ and $X\setminus U$, and observe that
\[ \sigma = \frac{e_B(U, X \setminus U)}{\card U \cdot \card{X\setminus U}} 
        \ge \frac{\mu\card X - \card U}{\card X - \card U} 
        \ge \mu - \frac{1}{k}.\]	
\end{quote}

The first equality above holds by definition, 
so it's remarkable that formalising the two subsequent inequalities 
took more than 70 proof lines to yield the weaker claim
$\sigma \ge \mu - 2/k$ (following Mehta).
The weakening doesn't matter because eventually we need to deduce $2\ell^{-5/12}\le \mu - 2/k$,
which holds for sufficiently large $\ell$ (thanks to $k\ge\ell$) and simply becomes part of the bigness predicate.
The first inequality above follows by careful counting of edges, while the second 
relies on $\card X \ge R(k,m)$ and properties of Ramsey numbers.

The proof continues as follows:

\begin{quote}
Let $S\subset U$ be a uniformly-chosen random subset of size~$b$, and let $Z = \card{N_B(S)\cap(X\setminus U)}$
be the number of common blue neighbours of $S$ in $X\setminus U$. By convexity, we have
\[ \mathbb{E}[Z] = \binom{m}{b}^{-1} \sum_{v\in X\setminus U}\binom{\card{N_B(v)\cap U}}{b} 
   \ge \binom{m}{b}^{-1} \binom{\sigma m}{b} \cdot \card{X\setminus U}. \]	
\end{quote}

The first equality requires a probabilistic argument. The expectation of $Z(S)$ for the randomly chosen subset $S$
can be formalised by setting up a probability space~\cite{edmonds-formal-probabilistic}.
``Convexity'' is doing a lot in the inequality. 
The function $f$ is said to be \textit{convex} if
$f \left (\sum_{i \in S} a_i y_i\right) \le \sum_{i \in S} a_i f (y_i))$
where $S$ is nonempty and finite, $a_i\ge0$ for $i\in S$ and the $a_i$ sum to~1.
In our proof, the $a_i = \card{X\setminus U}^{-1}$ and the convex function is 
the real-valued generalised binomial $\lambda x. \binom{x}{b}$, which unfortunately is not actually convex.
I adopted Mehta's elaborate fix for this problem, which was to define an alternative real-valued binomial
that was convex and agreed with the original where it mattered:
on the integers.

The proof continues:
\begin{quote}
Now \ldots{} recalling (10), and that $b = \ell^{1/4}$ and $m = \ell^{2/3}$, it follows that
\[ \mathbb{E}[Z] \ge \sigma^b \exp{\left(-\frac{b^2}{\sigma m}\right)} \cdot \card{X\setminus U} \ge \frac{\mu^b}{2} \cdot \card X,\]
and hence there exists a blue clique $S\subset U$ of size~b 
with at least this many common blue neighbours in $X\setminus U$, as required. $\square$
\end{quote}

Several dozen detailed and tedious lines of inequality calculations take care of this final part.
Lemma~4.1 is among the trickier theorems 
but its proof is typical of the sort of reasoning used throughout the paper: 
estimation and calculations involving edge density, edges and vertices.
On the other hand, the paper has only one further probabilistic argument: 
to derive a non-standard lower bound for Ramsey numbers.

\section{A progression of theorems}

It isn't possible to go through any other proofs in such detail as above, but let's have a glimpse at the main results of each section in the diagonal paper.
To begin, let's establish some conventions.
Recall that the book algorithm operates on state variables $X$, $Y$, $A$ and $B$.
Some other variables are defined in terms of them, notably $p_i$, the density of red edges between $X_i$ and~$Y_i$.

We have an issue with the subscripts:
the Isabelle and Lean formalisations both write $X_i$ for the value of $X$ after $i$ execution steps,
when Campos et al. \cite{campos-exponential-ramsey} write $X_{i-1}$.
Complicating matters further, everyone agrees to write $x_i$ 
for the value of the ``central vertex'' after $i$ execution steps.

In Isabelle, $X_i$ is written \isa{Xseq~i} and similarly for $Y$, $A$, $B$ and~$p_i$ .
It's important to note that $\epsilon=k^{-1/4}$, so $\epsilon$ depends on~$k$: 
it is formalised as \isa{eps~k} but within locales where $k$ is available
(\isa{No\_Cliques} and beyond) it's abbreviated to simply \isa{\isasymepsilon}.

A number of theorems refer to the situation after the algorithm halts.
The step number at which it halts is written \isa{halted\_point}.

\subsection{On red and density-boost steps} \label{sec:red}

The headline result of section 5 is entitled ``density-boost steps boost the density''.
This refers to the density of red edges between $X$ and $Y$: to be precise, the current values of those state variables.
The local definitions in the theorem statement correspond to abbreviations introduced by the authors.
The first assumption is that the current execution step is red or density-boost,
$i\in \cal R\cup\cal S$ in the diagonal paper.

\begin{isabelle}
\isacommand{lemma}\ Red\_5\_1:\isanewline
\ \ \isakeyword{assumes}\ i:\ "i\ \isasymin \ Step\_class\ \{red\_step,dboost\_step\}"\isanewline
\ \ \ \ \isakeyword{and}\ Big:\ "Big\_Red\_5\_1\ \isasymmu \ l"\isanewline
\ \ \isakeyword{defines}\ "p\ \isasymequiv \ pseq\ i"\isanewline
\ \ \isakeyword{defines}\ "x\ \isasymequiv \ cvx\ i"\isanewline
\ \ \isakeyword{defines}\ "X\ \isasymequiv \ Xseq\ i"\ \isakeyword{and}\ "Y\ \isasymequiv \ Yseq\ i"\isanewline
\ \ \isakeyword{defines}\ "NBX\ \isasymequiv \ Neighbours\ Blue\ x\ \isasyminter \ X"\isanewline
\ \ \isakeyword{defines}\ "NRX\ \isasymequiv \ Neighbours\ Red\ x\ \isasyminter \ X"\isanewline
\ \ \isakeyword{defines}\ "NRY\ \isasymequiv \ Neighbours\ Red\ x\ \isasyminter \ Y"\isanewline
\ \ \isakeyword{defines}\ "\isasymbeta \ \isasymequiv \ card\ NBX\ /\ card\ X"\isanewline
\ \ \isakeyword{shows}\ "red\_density\ NRX\ NRY\ \isasymge \ p\ -\ alpha\ (hgt\ p)\isanewline
\ \ \ \ \ \ \ \isasymor \ red\_density\ NBX\ NRY\ \isasymge \ p\ +\ (1-\isasymepsilon)\ *\ ((1-\isasymbeta )/\isasymbeta )\ *\ alpha(hgt\ p)\ \isasymand \ \isasymbeta \ >\ 0"
\end{isabelle}

One of the more interesting ``moments'' in the buildup to its proof was Lemma~5.6, 
where we find a little throwaway comment:
\begin{quote}
On the other hand, if we colour the edges independently at random, 
with probability $k^{-1/8}$  of being blue, then a simple (and standard) 1st moment argument shows that
\[ R(k, \ell^{3/4}) \ge \exp(c\, \ell^{3/4} \log k)\]
for some absolute constant $c > 0$.	
\end{quote}
I turned to Mehta's development to decipher this impossibly cryptic remark, 
arriving at a claim formalised in Isabelle as follows:
\begin{isabelle}
\isacommand{lemma}\ Ramsey\_number\_lower\_simple:\ \isanewline
\ \ \isakeyword{fixes}\ p::real\isanewline
\ \ \isakeyword{assumes}\ n:\ "n\isacharcircum k\ *\ p\ powr\ (k\isacharcircum 2\ /\ 4)\ +\ n\isacharcircum l\ *\ exp\ (-p\ *\ l\isacharcircum 2\ /\ 4)\ <\ 1"\isanewline
\ \ \isakeyword{assumes}\ p01:\ "0<p"\ "p<1"\ \isakeyword{and}\ "k>1"\ "l>1"\isanewline
\ \ \isakeyword{shows}\ "\isasymnot \ is\_Ramsey\_number\ k\ l\ n"
\end{isabelle}
This actually does have a fairly simple proof, 
appealing to a more general theorem about Ramsey numbers already in the library.%
\footnote{Thanks to a detailed proof written out for me by a colleague, Andrew Thomason.}

The actual claim ``density-boost steps boost the density'' is only evident in the following corollary of Lemma~5.1:

\begin{isabelle}
\isacommand{corollary}\ Red\_5\_2:\isanewline
\ \ \isakeyword{assumes}\ i:\ "i\ \isasymin \ Step\_class\ \{dboost\_step\}"\ \isanewline
\ \ \ \ \isakeyword{and}\ Big:\ "Big\_Red\_5\_1\ \isasymmu \ l"\isanewline
\ \ \isakeyword{shows}\ "pseq\ (Suc\ i)\ -\ pseq\ i\ \isasymge\isanewline
\ \ \ \ \ \ \ \ \ (1-\isasymepsilon)\ *\ ((1\ -\ beta\ i)\ /\ beta\ i)\ *\ alpha(hgt(pseq\ i))\ \isasymand \ beta\ i\ >\ 0"
\end{isabelle}

That the density indeed increases is more evident using
mathematical notation (adjusting for the Isabelle subscripting convention):
\[ p_{i+1}-p_i\ge (1-\epsilon)\left(\frac{1-\beta_i}{\beta_i}\right)\alpha_h. \]
Space prohibits saying much about~$\alpha_h$, 
but $\beta_i \in (0,1)$ is defined by
\[ \beta_i = \frac{\card{N_B(x_i)\cap X_i}}{\card{X_i}.} \]
The right-hand side above is strictly positive.

\subsection{Bounding the size of $Y$}

The book algorithm discards large numbers of vertices from $X$ and/or~$Y$
at each step, so we need to know that these sets don't get too small.
In particular, both density-boost steps ($\cal S$) and red steps ($\cal R$) discard from~$Y$.
Writing $s$ for $\card{\cal S}$ and $t$ for $\card{\cal R}$,
the headline result of section~6 is as follows:

\begin{lemma}[6.1]
	If $p_0=\Omega(1)$, then
	\[ \card Y \ge 2^{o(k)} p_0 ^ {s+t} \cdot \card {Y_0}. \]
\end{lemma}

The use of Landau symbols is perplexing. The actual meaning of $p_0=\Omega(1)$
is seemingly just that there is some positive floor $P_0$ on possible values of $p_0$,
specified in the formalisation by the locale \isa{P0\_min} presented in~\S\ref{sec:book}.
In Isabelle, $2^{o(k)}$ is formalised with reference to a specific $o(k)$ function:
$2k\log_2 (1 - 2 \sqrt{\epsilon} / P_0)$ turns out to serve the purpose if $P_0 > 2\sqrt{\epsilon}$.
Since $\epsilon=k^{-1/4}$, that condition holds for a large enough~$k$ and becomes part of the
``bigness assumption'' of this lemma.
An $o(k)$ function is one that grows more slowly than the identity; this particular function is actually negative.
As it is a function of $k$, not referring to the $k$ fixed in our locales, $\epsilon$ is formalised
as \isa{eps~k}.

\begin{isabelle}
\isacommand{definition}\ "ok\_fun\_61\ \isasymequiv \isanewline
\ \ \ \ \ \ \ \ \ \ \ \ \isasymlambda k.\ (2\ *\ real\ k\ /\ ln\ 2)\ *\ ln\ (1\ -\ 2\ *\ eps\ k\ powr\ (1/2)\ /\ p0\_min)"
\end{isabelle}

Now it is possible to formalise the statement of Lemma~6.1.

\begin{isabelle}
\isacommand{lemma}\ (\isakeyword{in}\ Book)\ Y\_6\_1:\isanewline
\ \ \isakeyword{assumes}\ big:\ "Big\_Y\_6\_1\ \isasymmu \ l"\isanewline
\ \ \isakeyword{defines}\ "st\ \isasymequiv \ Step\_class\ \{red\_step,dboost\_step\}"\isanewline
\ \ \isakeyword{shows}\ "card\ (Yseq\ halted\_point)\ /\ card\ Y0\ \isasymge\isanewline
\ \ \ \ \ \ \ \ \ \ 2\ powr\ (ok\_fun\_61\ k)\ *\ p0\ \isacharcircum \ card\ st"
\end{isabelle}

\subsection{Bounding the size of $X$} \label{sec:bounding-X}

As in the previous section, our task is to show that
density-boost ($\cal S$) and red ($\cal R$) steps don't cause too much damage. 
Recall that $\mu\in(0,1)$ is a parameter of the book algorithm and
that $s=\card{\cal S}$ and $t=\card{\cal R}$.
The headline result of section~7 refers to a further parameter $\beta\in(0,1)$, 
whose complicated definition is omitted here.

\begin{lemma}[7.1]
	\[ \card X \ge 2^{o(k)} \mu^\ell (1-\mu)^t \left(\frac{\beta}{\mu}\right) \card {X_0}. \]
\end{lemma}

And the same statement, formalised in Isabelle/HOL\@.

\begin{isabelle}
\isacommand{lemma}\ (\isakeyword{in}\ Book)\ X\_7\_1:\isanewline
\ \ \isakeyword{assumes}\ big:\ "Big\_X\_7\_1\ \isasymmu \ l"\isanewline
\ \ \isakeyword{defines}\ "\isasymD \ \isasymequiv \ Step\_class\ \{dreg\_step\}"\isanewline
\ \ \isakeyword{defines}\ "\isasymR \ \isasymequiv \ Step\_class\ \{red\_step\}"\ \isakeyword{and}\ "\isasymS \ \isasymequiv \ Step\_class\ \{dboost\_step\}"\isanewline
\ \ \isakeyword{shows}\ "card\ (Xseq\ halted\_point)\ \isasymge \ 2\ powr\ ok\_fun\_71\ \isasymmu \ k\ *\isanewline
\ \ \ \ \ \ \ \ \ \ \ \ \ \ \ \ \isasymmu \isacharcircum l\ *\ (1-\isasymmu )\ \isacharcircum \ card\ \isasymR \ *\ (bigbeta/\isasymmu )\ \isacharcircum \ card\ \isasymS \ *\ card\ X0"
\end{isabelle}

In this case, the construction of the $o(k)$ function is less straightforward.
Lemma~7.1 follows from a succession of other lemmas, such as Lemma~7.2:
\[ \prod{i\in\cal R}\, \frac{\card{X_{i+1}}}{\card{X_i}} \ge 2^{o(k)} (1-\mu)^t. \]

So \isa{ok\_fun\_71}, the $o(k)$ function for Lemma~7.1, is defined as 
the sum of four other $o(k)$ functions used in the other lemmas building up to it.%
\footnote{Lemma $n$.1 is generally the final result of section $n$, proved using $n$.2, $n$.3, etc. But not if $n=8$.}
This section, all six pages of it, was painful to formalise.

\subsection{The zigzag lemma}

This lemma bounds the number $t$ of density-boost steps, limiting their wastage of vertices in~$X$.
It refers to the central vertex~$x_i$ and to $\cal S^*$, a subset of the density-boost steps 
with a highly technical definition.
Now Lemma~8.1 is stated as follows:
\[ \sum_{i\in\cal S^*} \frac{1-\beta_i}{\beta_i} \le t + o(k). \]

In the Isabelle version, the $o(k)$ function is written explicitly as $k^{19/20}$:

\begin{isabelle}
\isacommand{lemma}\ (\isakeyword{in}\ Book)\ ZZ\_8\_1:\isanewline
\ \ \isakeyword{assumes}\ big:\ "Big\_ZZ\_8\_1\ \isasymmu \ l"\ \isanewline
\ \ \isakeyword{defines}\ "\isasymR \ \isasymequiv \ Step\_class\ \{red\_step\}"\isanewline
\ \ \isakeyword{defines}\ "sum\_SS\ \isasymequiv \ (\isasymSum i\isasymin dboost\_star.\ (1\ -\ beta\ i)\ /\ beta\ i)"\isanewline
\ \ \isakeyword{shows}\ "sum\_SS\ \isasymle \ card\ \isasymR\ +\ k\ powr\ (19/20)"
\end{isabelle}

The section is only three pages long and yielded one of the shortest formalisations of the entire development. 

\subsection{An exponential improvement far from the diagonal}

Sections 9--12 of the diagonal paper use the accumulated knowledge about the book algorithm 
to achieve significant new results.
``Far from the diagonal'' means $\ell\le k/9$.
The headline result of section~9 is stated as follows. 

\begin{theorem}[9.1]
If $\gamma\le1/10$ and $\delta=\gamma/20$, then
	\[ R(k,\ell)\le e^{-\delta k + o(k)} \binom{k+\ell}{\ell} \]
for all $k$, $\ell\in \mathbb{N}$ with $\gamma=\frac{\ell}{k + \ell}$.
\end{theorem}

The formal version below may look significantly weaker, imposing a
``for all big enough $\ell$'' condition not present in the original.
It also assumes that \isa{p0\_min}, the minimum possible $p_0$, must not exceed approximately $0.89$.
On the plus side, the $o(k)$ function has been replaced by simply~1.

\begin{isabelle}
\isacommand{theorem}\ Far\_9\_1:\isanewline
\ \ \isakeyword{fixes}\ l\ k::nat\isanewline
\ \ \isakeyword{fixes}\ \isasymdelta \ \isasymgamma ::real\isanewline
\ \ \isakeyword{defines}\ "\isasymgamma \ \isasymequiv \ real\ l\ /\ (real\ k\ +\ real\ l)"\isanewline
\ \ \isakeyword{defines}\ "\isasymdelta \ \isasymequiv \ \isasymgamma /20"\isanewline
\ \ \isakeyword{assumes}\ \isasymgamma :\ "\isasymgamma \ \isasymle \ 1/10"\ \isanewline
\ \ \isakeyword{assumes}\ big:\ "Big\_Far\_9\_1\ \isasymgamma \ l"\isanewline
\ \ \isakeyword{assumes}\ p0\_min\_91:\ "p0\_min\ \isasymle \ 1\ -\ (1/10)\ *\ (1\ +\ 1/15)"\isanewline
\ \ \isakeyword{shows}\ "RN\ k\ l\ \isasymle \ exp\ (-\isasymdelta *k\ +\ 1)\ *\ (k+l\ choose\ l)"
\end{isabelle}

Also notable is that all the lemmas presented before now were proved
within the \isa{Book} locale, providing the full environment and assumptions of the book algorithm.
Theorem 9.1 is proved in locale \isa{P0\_min}, which specifies nothing other than \isa{p0\_min}.

It will be instructive to see how the pieces fit together. Theorem~9.1 requires the following lemma:

\begin{lemma}[9.2]
  Let $\gamma\le1/10$, and let $k$, $\ell\in\mathbb{N}$ be sufficiently large integers with $\gamma=\frac{\ell}{k + \ell}$.
  Let $\delta=\gamma/20$ and $\eta\le \gamma/15$, and suppose that
	\[ n \ge e^{-\delta k} \binom{k+\ell}{\ell}. \]
	Then every red-blue colouring of the edges of the complete $n$-graph in which the density of red edges is at least
	$1-\gamma-\eta$ contain either a red $k$-clique or a blue $\ell$-clique.
\end{lemma}

The proof relies on the lemmas proved earlier within the locale \isa{Book}:
they refer to the book algorithm
and its implicit assumption that there exists \textbf{no} red $k$-clique or blue $\ell$-clique,
so we must replace the conclusion above by falsity.
We formalise the lemma in a series of steps.
First, we prove a modified form of Lemma~9.2 under the assumptions of the book algorithm%
\footnote{Recall that locale \texttt{\slshape Book'} is like \texttt{\slshape Book} but replacing the parameter~$\mu$ by $\gamma=\ell/(k+\ell)$.}
and concluding \isa{False}.

\begin{isabelle}
\isacommand{lemma}\ (\isakeyword{in}\ Book')\ Far\_9\_2\_aux:\isanewline
\ \ \isakeyword{fixes}\ \isasymdelta \ \isasymeta ::real\isanewline
\ \ \isakeyword{defines}\ "\isasymdelta \ \isasymequiv \ \isasymgamma /20"\isanewline
\ \ \isakeyword{assumes}\ "real\ (card\ X0)\ \isasymge \ nV/2"\ "card\ Y0\ \isasymge \ nV\ div\ 2"\ "p0\ \isasymge \ 1-\isasymgamma -\isasymeta "\isanewline
\ \ \isakeyword{assumes}\ "\isasymgamma \ \isasymle \ 1/10"\ \isakeyword{and}\ \isasymeta :\ "0\isasymle \isasymeta "\ "\isasymeta \ \isasymle \ \isasymgamma /15"\isanewline
\ \ \isakeyword{assumes}\ "real\ nV\ \isasymge \ exp\ (-\isasymdelta \ *\ k)\ *\ (k+l\ choose\ l)"\ \isanewline
\ \ \isakeyword{assumes}\ big:\ "Big\_Far\_9\_2\ \isasymgamma \ l"\isanewline
\ \ \isakeyword{shows}\ False
\end{isabelle}

Its proof formalises that of Lemma~9.2 in the diagonal paper. 
But a little more work is needed to get the lemma statement into the correct form:
specifically, to set up the invocation of the book algorithm given a complete graph
with a red-blue colouring of its edges having no red $k$-clique or blue $\ell$-clique.
Assuming the density of red edges is large enough and that $\mu\in(0,1)$, 
it is straightforward to partition the vertex set into $X_0$ and $Y_0$ with a sufficiently large red-edge density between them.
The conclusion is expressed using the \isa{Book} predicate, 
so this lemma makes it easy to move up from the
\isa{No\_Cliques} locale to the \isa{Book} locale.

\begin{isabelle}
\isacommand{lemma}\ (\isakeyword{in}\ No\_Cliques)\ to\_Book:\isanewline
\ \ \isakeyword{assumes}\ "p0\_min\ \isasymle \ graph\_density\ Red"\isanewline
\ \ \isakeyword{assumes}\ "0\ <\ \isasymmu "\ "\isasymmu \ <\ 1"\isanewline
\ \ \isakeyword{obtains}\ X0\ Y0\ \isakeyword{where}\ "l\isasymge 2"\ "card\ X0\ \isasymge \ real\ nV\ /\ 2"\ "card\ Y0\ =\ gorder\ div\ 2"\isanewline
\ \ \ \ \isakeyword{and}\ "X0\ =\ V\ \isasymsetminus \ Y0"\ "Y0\isasymsubseteq V"\ \isanewline
\ \ \ \ \isakeyword{and}\ "graph\_density\ Red\ \isasymle \ gen\_density\ Red\ X0\ Y0"\isanewline
\ \ \ \ \isakeyword{and}\ "Book\ V\ E\ p0\_min\ Red\ Blue\ l\ k\ \isasymmu \ X0\ Y0"
\end{isabelle}
Note that \isakeyword{obtains} is a convenient way of expressing an existential conclusion.
Here it is logically equivalent to $\exists X_0 Y_0. \, \ldots$.
A corollary (similar and omitted) is \isa{to\_Book'}, 
for instantiating the \isa{Book'} locale.

And so we obtain Lemma~9.2 in its final form, proved in the \isa{No\_Cliques} locale. This manipulation of locales is not difficult
and provides a uniform treatment of the framework of the book algorithm and associated notation.

\begin{isabelle}
\isacommand{lemma}\ (\isakeyword{in}\ No\_Cliques)\ Far\_9\_2:\isanewline
\ \ \isakeyword{fixes}\ \isasymdelta \ \isasymgamma \ \isasymeta ::real\isanewline
\ \ \isakeyword{defines}\ "\isasymgamma \ \isasymequiv \ l\ /\ (real\ k\ +\ real\ l)"\isanewline
\ \ \isakeyword{defines}\ "\isasymdelta \ \isasymequiv \ \isasymgamma /20"\isanewline
\ \ \isakeyword{assumes}\ "graph\_density\ Red\ \isasymge \ 1-\isasymgamma -\isasymeta "\ \isakeyword{and}\ "p0\_min\ \isasymle \ 1-\isasymgamma -\isasymeta "\ \ \isanewline
\ \ \isakeyword{assumes}\ "\isasymgamma \ \isasymle \ 1/10"\ \isakeyword{and}\ "0\isasymle \isasymeta "\ "\isasymeta \ \isasymle \ \isasymgamma /15"\isanewline
\ \ \isakeyword{assumes}\ "real\ nV\ \isasymge \ exp\ (-\isasymdelta \ *\ k)\ *\ (k+l\ choose\ l)"\ \isanewline
\ \ \isakeyword{assumes}\ big:\ "Big\_Far\_9\_2\ \isasymgamma \ l"\isanewline
\ \ \isakeyword{shows}\ False
\end{isabelle}

From this, formalising the proof of Theorem~9.1 takes another 450+ lines.
The proof is by contradiction, assuming $R(k,\ell) > e^{-\delta k + o(k)} \binom{k+\ell}{\ell}$
and defining $n$ (the number of vertices in the graph) to be $R(k,\ell)-1$.
We immediately obtain a clique-free red/blue colouring of the edges, and eventually
use Lemma~9.2 (and therefore the book algorithm) in one case, for this graph restricted to a subset $U$ of the vertices.
There is a subtle error in the proof of 9.1 that I was able to patch with the help of Simon Griffiths, one of the authors.

\subsection{An exponential improvement a little closer to the diagonal}

For section~10, ``a little closer to the diagonal'' means $\ell\le k/4$,
and the headline result is stated as follows. 

\begin{theorem}[10.1]
If $\gamma\le1/5$ and $\delta=\gamma/40$, then
	\[ R(k,\ell)\le e^{-\delta k + o(k)} \binom{k+\ell}{\ell} \]
for all $k$, $\ell\in \mathbb{N}$ with $\gamma=\frac{\ell}{k + \ell}$.
\end{theorem}

The similarity to the previous theorem should be obvious, and it is proved in a somewhat similar fashion from the following lemma.

\begin{lemma}[10.2]
  Let $k$, $\ell\in\mathbb{N}$ be sufficiently large, and set $\gamma=\frac{\ell}{k + \ell}$.
  If $1/10 \le \gamma\le 1/5$ and
	\[ n \ge e^{-k/200} \binom{k+\ell}{\ell}, \]
	then every red-blue colouring of the edges of the complete $n$-graph in which the density of red edges is at least
	$1-\gamma$ contain either a red $k$-clique or a blue $\ell$-clique.
\end{lemma}

The proof of the theorem from the lemma has much in common with the analogous proof for section~9 and even an analogous error, 
but some differences are worth noting.
Here is the formalisation of the headline result:

\begin{isabelle}
\isacommand{theorem}\ Closer\_10\_1:\isanewline
\ \ \isakeyword{fixes}\ l\ k::nat\isanewline
\ \ \isakeyword{fixes}\ \isasymdelta \ \isasymgamma ::real\isanewline
\ \ \isakeyword{defines}\ "\isasymgamma \ \isasymequiv \ real\ l\ /\ (real\ k\ +\ real\ l)"\isanewline
\ \ \isakeyword{defines}\ "\isasymdelta \ \isasymequiv \ \isasymgamma /40"\isanewline
\ \ \isakeyword{defines}\ "\isasymgamma 0\ \isasymequiv \ min\ \isasymgamma \ (0.07)"\isanewline
\ \ \isakeyword{assumes}\ big:\ "Big\_Closer\_10\_1\ \isasymgamma 0\ l"\isanewline
\ \ \isakeyword{assumes}\ \isasymgamma :\ "\isasymgamma \ \isasymle \ 1/5"\ \isanewline
\ \ \isakeyword{assumes}\ "p0\_min\ \isasymle \ 1\ -\ 1/5"\isanewline
\ \ \isakeyword{shows}\ "RN\ k\ l\ \isasymle \ exp\ (-\isasymdelta *k\ +\ 3)\ *\ (k+l\ choose\ l)"
\end{isabelle}

As before, it looks weaker than it should be. Once again it assumes, contrary to the statement of 10.1,
that $\ell$ needs to be sufficiently large --- and worse, the relevant ``bigness lemma''
refers to the mysterious parameter $\gamma_0$, defined to be $\min(\gamma,0.07)$.
In fact $0.07 < 1/10 - 1/36$ and is concerned with an application of Theorem~9.1 (when $\gamma\le 1/10$)
when $\ell$ has been replaced by a somewhat smaller value arising from the restriction of our graph to a subgraph.
It turns out we can deduce $k\ge36$ from our bigness assumptions (I suspect that it must be vastly bigger).
The other point is that once again the claimed $o(k)$ function is degenerate, in this case simply the value~3.

An unconditional version of Theorem~10.1 seems necessary.
We can obtain one by defining a suitable $o(k)$ function.
That will be the function that returns~3
if the given~$k$ is big enough according to the theorem and otherwise returns $\delta k$,
due to the long-established inequality $R(k,\ell)\le \binom{k+\ell}{\ell}$.
This version still has a stray but harmless assumption about \isa{p0\_min}.

\begin{isabelle}
\isacommand{theorem}\ Closer\_10\_1\_unconditional:\isanewline
\ \ \isakeyword{fixes}\ l\ k::nat\isanewline
\ \ \isakeyword{fixes}\ \isasymdelta \ \isasymgamma ::real\isanewline
\ \ \isakeyword{defines}\ "\isasymgamma \ \isasymequiv \ real\ l\ /\ (real\ k\ +\ real\ l)"\isanewline
\ \ \isakeyword{defines}\ "\isasymdelta \ \isasymequiv \ \isasymgamma /40"\isanewline
\ \ \isakeyword{assumes}\ \isasymgamma :\ "0\ <\ \isasymgamma "\ "\isasymgamma \ \isasymle \ 1/5"\ \isanewline
\ \ \isakeyword{assumes}\ p0\_min\_101:\ "p0\_min\ \isasymle \ 1\ -\ 1/5"\isanewline
\ \ \isakeyword{shows}\ "RN\ k\ l\ \isasymle \ exp\ (-\isasymdelta *k\ +\ ok\_fun\_10\_1\ \isasymgamma \ k)\ *\ (k+l\ choose\ l)"
\end{isabelle}

\subsection{Concluding the proof of the main theorem} \label{sec:main}

The proof ends in a final blaze of technical complexity, of which only hints can be shown.
Section~11 defines two functions, for $x$, $y\in (0,1)$ and (eventually) $\mu=2/5$:
\begin{align*}
	F(x,y) &= \frac{1}{k} \log_2 R(k, k-xk)+x+y \\
	G_\mu(x,y) &= \log_2 \left(\frac{1}{\mu}\right) + x \cdot \log_2\left(\frac{1}{1-\mu}\right)
		           + y \cdot \log_2\left(\frac{\mu(x+y)}{y}\right).
\end{align*}

The objective of section~11, which is entitled ``from off-diagonal to diagonal'', is this:

\begin{theorem}[11.1]
  Fix $\mu\in(0,1)$ and $\eta>0$. We have
  	\[ \frac{\log_2 R(k)}{k} \le 
  		\max_{{0\le x\le 1} \atop {0\le y \le \mu x / (1-\mu)+\eta}} 
  		\min \bigl\{ F(x,y), G_\mu(x,y)\bigr\} + \eta\]
	for all sufficiently large $k\in\mathbb N$.
\end{theorem}

It seemed that we are nearing our destination, because an upper bound on $\log_2 R(k)/k$
looks like giving an upper bound on~$R(k)$.
Unfortunately, this doesn't quite work because $F(x,y)$ hides a dependence on~$k$ that shouldn't be there.
The correct version of Theorem~11.1 is stated in terms of other functions,
which are actually defined in section~12.
\begin{align*}
	h(p) &= -p \log_2 p - (1-p) \log_2(1-p) \\
	f_1(x,y) &= x + y + (2-x)\cdot h\left(\frac{1}{2-x}\right) \\
	f_2(x,y) &= f_1(x,y) - \frac{\log_2 e}{40}\left(\frac{1-x}{2-x}\right) \\
	f(x,y) &= \begin{cases}
		f_1(x,y) & \text{if $x<3/4$}\\ f_2(x,y) & \text{if $x\ge 3/4$}
	\end{cases}
\end{align*}

The definitions above are formalised in the obvious way, where \isa{ff} is $f$ and \isa{GG} is $G$.
But we seemingly need to modify the theorem statement by imposing a lower bound of~1.9 on the minimum:
\begin{isabelle}
\isacommand{definition}\ "ffGG\ \isasymequiv \ \isasymlambda \isasymmu \ x\ y.\ max\ 1.9\ (min\ (ff\ x\ y)\ (GG\ \isasymmu \ x\ y))"
\end{isabelle}

Here is Theorem~11.1 as formalised. There are several modifications, all making it weaker than the original version.
\begin{isabelle}
\isacommand{theorem}\ (\isakeyword{in}\ P0\_min)\ From\_11\_1:\isanewline
\ \ \isakeyword{assumes}\ \isasymmu :\ "0\ <\ \isasymmu "\ "\isasymmu \ \isasymle \ 2/5"\ \isakeyword{and}\ "0\ <\ \isasymeta"\ "\isasymeta \ \isasymle \ 1/12"\isanewline
\ \ \ \ \isakeyword{and}\ p0\_min12:\ "p0\_min\ \isasymle \ 1/2"\ \isakeyword{and}\ big:\ "Big\_From\_11\_1\ \isasymeta \ \isasymmu \ k"\isanewline
\ \ \isakeyword{shows}\ "log\ 2\ (RN\ k\ k)\ /\ k\ \isasymle\isanewline
\ \ \ \ \ \ \ \ \ \ \ \ (SUP\ x\ \isasymin \ \{0..1\}.\ SUP\ y\ \isasymin \ \{0..3/4\}.\ ffGG\ \isasymmu \ x\ y\ +\ \isasymeta )"
\end{isabelle}

Note that this theorem is situated in the trivial locale \isa{P0\_min}.
The proof, somewhat like others we have discussed, begins by setting $n=R(k,k)-1$,
obtaining a clique-free red/blue colouring of the edges of the complete $n$-graph.
There is a case analysis on whether the density of red or blue edges is greater;
either way, the \isa{Book} locale is interpreted, allowing the use of lemmas proved earlier.

We are finally close to our objective of proving
\begin{theorem}[1.1]
  There exists $\epsilon > 0$ such that
  	\[ R(k)\le(4-\epsilon)^k \]
	for all sufficiently large $k\in\mathbb N$.
\end{theorem}

The nasty sting in the tail comes, appropriately, at the very end of the argument.
The formal version of this theorem is similar but uses the function \isa{ffGG},
as in the formalisation of Theorem~11.1.

\begin{lemma}[12.3]
If $\delta \le 2^{-11}$, then 
	\[ \max_{{0\le x\le 1} \atop {0\le y \le 3/4}} 
  		\min \bigl\{ f(x,y), g(x,y)\bigr\} < 2-\delta. \]
\end{lemma}

The authors give a full two-page proof of this claim,
illustrated by a diagram highlighting the regions where either $f$ or $g$ is too large.%
\footnote{The function $g$ is simply $G_{2/5}$. And incidentally, $h$ is the binary entropy function.}
We are fine if the two regions don't overlap, 
but they are separated only by a hair's breadth.
The authors remark that the proof is ``a simple calculation'' that could be checked by computer,
which takes us to the interesting topic of verified calculations.

\section{Formal reasoning about computer algebra}

During the 1990s, many researchers sought to combine the capabilities of computer algebra systems
(such as Maple and Mathematica) with interactive theorem provers.
Computer algebra systems could deliver incorrect answers, 
due to overly aggressive simplification or outright bugs.
A student of mine, Clemens Ballarin, worked on this problem~\cite{ballarin-pragmatic}.
A difficulty was finding an application,
since most computer algebra users are not performing proofs.
But the diagonal Ramsey proof needs computer algebra everywhere: derivatives (typically for identifying maxima),
limits, continuity, Landau symbols, high precision exact real arithmetic.
We are frequently forced to perform precise numerical calculations 
or to prove the existence of a large enough $\ell$ satisfying some complicated condition.

Isabelle/HOL has a number of automatic built-in tools to carry out such tasks:
\begin{itemize}
	\item Eberl's \textbf{real\_asymp} proof method~\cite{eberl-verified-real}, to prove theorems about limits and Landau symbols for real-valued functions 
	\item Hölzl's \textbf{approximation} proof method~\cite{hoelzl-inequalities}, to perform exact real calculations using interval arithmetic
	\item the rule set \isa{derivative\_eq\_intros}, for performing symbolic differentiation
\end{itemize}

The last item needs further explanation, because it is not packaged as a neat proof method.
Taking the derivative of a function by the repeated application of syntactic rules is an easy programming exercise
dating back nearly 60 years~\cite{weissman-lisp-primer}.
Theorems stating the rules for differentiating functions such as multiplication, division, sine or cosine,
packaged in the identifier \isa{derivative\_eq\_intros}, can take derivatives of quite large expressions
when combined with a call to the simplifier.

Here is an example, from Lemma~9.3, 
proving that the function $1-1/{200x}$ is concave on $[1/10,1/5]$
by calculating its second derivative and noting that $-1/(100x^3)\leq0$ if $x\in(0,1)$. 
\begin{isabelle}
\isacommand{show}\ "concave\_on\ \{1/10..1/5\}\ (\isasymlambda x.\ 1\ -\ 1/(200*x))"\isanewline
\isacommand{proof}\ (intro\ f''\_le0\_imp\_concave)\isanewline
\ \ \isacommand{fix}\ x::real\isanewline
\ \ \isacommand{assume}\ "x\ \isasymin \ \{1/10..1/5\}"\isanewline
\ \ \isacommand{then}\ \isacommand{have}\ x01:\ "0<x"\ "x<1"\ \isacommand{by}\ auto\isanewline
\ \ \isacommand{show}\ "((\isasymlambda x.\ (1\ -\ 1/(200*x)))\ has\_real\_derivative\ 1/(200*x\isacharcircum 2))\ (at\ x)"\isanewline
\ \ \ \ \isacommand{using}\ x01\ \isacommand{by}\ (intro\ derivative\_eq\_intros\ |\ force\ simp:\ eval\_nat\_numeral)+\isanewline
\ \ \isacommand{show}\ "((\isasymlambda x.\ 1/(200*x\isacharcircum 2))\ has\_real\_derivative\ -1/(100*x\isacharcircum 3))\ (at\ x)"\isanewline
\ \ \ \ \isacommand{using}\ x01\ \isacommand{by}\ (intro\ derivative\_eq\_intros\ |\ force\ simp:\ eval\_nat\_numeral)+\isanewline
\ \ \isacommand{show}\ "-1/(100*x\isacharcircum 3)\ \isasymle \ 0"\isanewline
\ \ \ \ \isacommand{using}\ x01\ \isacommand{by}\ (simp\ add:\ divide\_simps)\isanewline
\isacommand{qed}\ auto
\end{isabelle}

The diagonal Ramsey development includes 16 symbolic differentiations, 30 calls to \textbf{approximation} and 60 calls to \textbf{real\_asymp}.
Some of the latter would be difficult to prove without those tools.
Let's look at the formal proof that
\[ \frac{k}{\log 2} \cdot \log\left(1-\frac{1}{k(1-\mu)}\right) = o(k) \]

This is the $o(k)$ function mentioned in the statement of Lemma 7.2, which we saw in \S\ref{sec:bounding-X} above.
The formal proof requires merely unfolding the definition, then calling \textbf{real\_asymp}.%
\footnote{The complexity class $o(k)$ is formalised as \texttt{\slshape o(real)}, where \texttt{\slshape real} is the injection from $\mathbb N$ into $\mathbb R$.}

\begin{isabelle}
\isacommand{definition}\ "ok\_fun\_72\ \isasymequiv \ \isasymlambda \isasymmu \ k.\ (real\ k\ /\ ln\ 2)\ *\ ln\ (1\ -\ 1\ /\ (k\ *\ (1-\isasymmu )))"\ \isanewline
\isacommand{lemma}\ ok\_fun\_72:\isanewline
\ \ \isakeyword{assumes}\ "\isasymmu <1"\ \isanewline
\ \ \isakeyword{shows}\ "ok\_fun\_72\ \isasymmu \ \isasymin \ o(real)"\isanewline
\isacommand{using}\ assms\ \isacommand{unfolding}\ ok\_fun\_72\_def\ \isacommand{by}\ real\_asymp
\end{isabelle}

Many of the asymptotic lemmas claim that some property $P(\mu,\ell)$ holds for all sufficiently large~$\ell$ 
and all $\mu$ ranging over some closed subinterval $[\mu_0,\mu_1]\subset(0,1)$.
The \textbf{real\_asymp} method does not work for intervals, but if $P(\mu,\ell)$ is preserved as $\mu$ increases,
then it is enough to check $P(\mu_0)$, and if it is preserved as $\mu$ decreases, it is enough to check $P(\mu_1)$.

The calls to \textbf{approximation} come in two forms. Most are simple calculations, such as 
$835.81 \le  3 \cdot \log_2 (5/3) / 5 \cdot \log_2 2727$, which requires high precision.
However, the proof of Lemma~12.3, mentioned above, twice requires reasoning over a closed interval.
\begin{isabelle}
\ \ \isacommand{have}\ ?thesis\ \isakeyword{if}\ "y\ \isasymin \ \{1/10\ ..\ 3/4\}"\isanewline
\ \ \ \ \isacommand{using}\ that\ \isacommand{unfolding}\ gg\_eq\ x\_of\_def\ \isanewline
\ \ \ \ \isacommand{by}\ (approximation\ 24\ splitting:\ y\ =\ 12)
\end{isabelle}

The \isa{splitting} keyword controls the interval splitting in the search. The claim being proved here is
\begin{alltt}
   log 2 (5/2) + (3 * y / 5 + 5454 / 10^4) * log 2 (5/3) +
   y * log 2 (2 * (3 * y / 5 + 5454 / 10^4 + y) / (5 * y))
   \(\le\) 2 - 1/2^11
\end{alltt}

The two \textbf{approximation} calls with splitting take a total of nearly 30 seconds to execute.
By way of comparison, the entire diagonal Ramsey proof (including those calls) run in 519 seconds.

\section{Discussion and conclusions}

This project was a huge amount of work. The formalisation took 251 days, just over eight months, and I often needed advice from others.
The de Bruijn index of this development (the ratio between the formal and informal text, after compression)
is relatively high, at~5.6, a sign of the work's difficulty. (Common is 4 \cite{edmonds-formalising-regularity}.)

Mehta's Lean~3 proof%
\footnote{\url{https://github.com/b-mehta/exponential-ramsey/blob/main/src/main_results.lean}}
was valuable in several respects: to resolve errors and ambiguities in the paper, to suggest lemmas, 
to give clues as to difficulty of a particular proof
and simply to demonstrate that the work could be formalised at all.
My chief borrowings are as follows:
\begin{itemize}
	\item the convexity argument in Lemma~4.1, as discussed in~\S\ref{sec:peek}
	\item an unusual Ramsey lower bound for Lemma~5.6, as discussed in~\S\ref{sec:red}
	\item the solution to the hidden dependence on~$k$ in $F(x,y)$, as discussed in~\S\ref{sec:main}
\end{itemize}
Other borrowings are indicated by comments in my formalisation~\cite{Diagonal_Ramsey-AFP},
which otherwise stands apart from his.

What do we learn from this project? Mainly, it provides more evidence that proof assistants 
--- in particular Isabelle/HOL --- can cope with new mathematics.
Although the techniques used hardly go beyond undergraduate mathematics, 
it is an intricate construction
combining probability, graph theory, asymptotics and high precision numerical calculations.
It is the sort of proof that would surely attract doubters, but with two formalisations, 
in different systems, there can be little room for doubt anymore.
Machine proofs inspire confidence because they are precise, explicit and checkable.
More to the point: anyone who has experienced the difficulty of getting
a proof assistant to accept anything will be confident when it  
acccepts your claim as a theorem.

With so much mathematics formalised, both in both Lean and in Isabelle~\cite{paulson-large-scale}, 
it's reasonable to say that the point has been made and that we can now stop.
But still: so-called proof assistants should actually be assisting mathematicians 
in their work, as they already assist computer scientists trying to prove the correctness 
of, for example, distributed systems~\cite{gomes-verifying-strong}.

One possible line of attack would begin with the Isabelle proof just finished.
The diagonal paper is full of seemingly arbitrary numerical choices.
The Isabelle proof is a living document that you can easily tinker with, 
making a change and immediately discovering what goes wrong.
Somebody who knows what they are doing could almost certainly find
a sharper bound or a simpler proof.

\paragraph*{Acknowledgements}
I'm grateful to Mantas Baksys, Simon Griffiths, Bhavik Mehta and Andrew Thomason for mathematical advice.
Simon provided a detailed plan for dealing with an error in the proofs of lemmas 9.1 and~10.1, 
while Andrew sketched out the standard probabilistic proof of the lower bound for Ramsey numbers, 
which remarkably did not appear to be written up anywhere.
Help with specialised Isabelle proof methods was provided by
Manuel Eberl (\textbf{real\_asymp}) and Fabian Immler (\textbf{approximation}).
Sophie Tourret read a draft of the paper and provided detailed, thoughtful comments.

\begin{flushleft}
\bibliographystyle{plain}
\bibliography{string,atp,general,isabelle,theory,crossref}
\end{flushleft}

\end{document}